\documentclass[12pt]{article}

\usepackage{graphicx}

\title{Origin of  Correlations between \\
Central Black Holes Masses and Galactic Bulge Velocity Dispersions}

\author{V. I. Dokuchaev\thanks{dokuchaev@inr.npd.ac.ru},
        ~Yu. N. Eroshenko\thanks{erosh@ns.ufn.ru} \\
{\small\sl Institute for Nuclear Research of the Russian Academy of
Sciences, Moscow}}

\begin{document}
\date{}
\maketitle

\begin{abstract}
We argue that the observed correlations between central black holes
masses $M_{BH}$ and galactic bulge velocity dispersions $\sigma_e$ in the
form $M_{BH}\propto\sigma_e^4$ may witness on the pregalactic origin of
massive black holes. Primordial black holes  would be the centers for
growing protogalaxies which experienced multiple mergers with ordinary
galaxies. This process is accompanied by the merging of black holes in
the galactic nuclei.
\end{abstract}

Recent observations$^1$ demonstrate that no less than 20\%
of regular galaxies
contain the supermassive black holes (SMBH) in their nuclei.
There are observed  correlations between the mass of the
central SMBH $M_{BH}$ in the galactic nucleus and
velocity dispersion $\sigma_e$ at the bulge half-optical-radius$^2$:
\begin{equation}
M_{BH}=1.2(\pm0.2)\times10^8\left(
\frac{\sigma_e}{200\mbox{~km/s}}\right)
^{3.75(\pm0.3)}M_{\odot}.\label{korsig}
\end{equation}
We use term ``bulge'' as for elliptical
galaxies and for central spheroidal parts of spiral galaxies.
Note what another study$^3$ which
based on poorer observational data
gives the correlation in the different form
$M_{BH}\propto\sigma_e^{4.8(\pm0.5)}$.

The origin of discussed correlations is quite uncertain. The simplest
assumption that the growth of the SMBH mass depends on the bulge processes
meets with the problem of different scales: the galactic bulge  scale is a
few kpc, whereas the linear scale
of accretion disk around of SMBH is much less than $1$~pc.
Some deterministic
mechanism$^4$ is needed for huge mass transfer from bulge to its innermost part.
In these scenarios$^6$ the
SMBHs are formed deep inside of the gravitational potential well of the
galactic or protogalalctic nuclei.

Here we explore the alternative approach by supposing that observed
correlations are stochastic in origin. Our basic assumption is the
existence in the Universe of pre-galactic population of black holes with
masses $M_h\sim10^5M_{\odot}$ before the recombination time. The similar
hypothesis of the existence of pre-galactic massive black hole population
was used by Fukugita and Turner$^5$ for interpretation of quasar
evolution. The specific possibility$^{7,8}$ is the formation of
primordial black holes (PBH) in the early Universe, or the formation of
pre-galactic black holes during cosmological phase transitions$^9$.

The supposed PBHs are mixed with dark matter due to there
cosmological origin. So the total mass of these PBHs in any galaxy $\sum M_h$
would be proportional to galactic dark matter halo mass $M$. As a result the
correlation $\sum M_h\propto M$ is primary in this model and
the
aforementioned observed correlations $M_{BH}\propto \sigma_e
^4$ would be secondary and approximate
in origin due to complicated process of galactic formation.
We may clarify this thesis as follows.

Cosmological fluctuation
power spectrum $P(k)$ in the confined mass region can be
approximated by the power law with the effective index $n=d\ln P(k)/d\ln
k$.
The effective galactic mass (mass of $L^*$ galaxy)
formed at red-shift $z$ is$^{10}$
\begin{equation}
M=M_0(1+z)^{-\frac{6}{n+3}},
\end{equation}
where $M_0=const$. On galactic mass scales
$M=(10^{10}\div10^{12})M_{\odot}$ value of $n$ vary from $-2.28$ to
$-1.98$ and $M_0$ vary from $2.5\cdot10^{16}M_{\odot}$ to
$7\cdot10^{14}M_{\odot}$ respectively. Velocity dispersion is estimated as
$\sigma_e^2\simeq GM/R$, where
$R=\left(3M/4\pi\kappa\rho(z)\right)^{1/3}$ and $\rho(z)=\rho_0(1+z)^3$,
$\kappa=178$, $\rho_0$ is the current density of cold dark matter.

For PBH cosmological density parameter $\Omega_h$ and effective PBH merging
in the galaxies the preceding relations gives the final mass of the central
SMBHs
\begin{equation}
M_{BH}=\psi\Omega_hM=\psi\Omega_{h}\sigma_e^{\frac{12}{1-n}}
M_0^{-\frac{n+3}{1-n}}\left(4\pi G^3\rho_0/3\right)^{-\frac{2}{1-n}},
\label{bhn}
\end{equation}
where factor $\psi$ is responsible for the possible additional growth of
the central SMBH due to accretion. From (\ref{bhn}) it follows
\begin{equation}
M_{BH}=(0.91\div1.03)\cdot10^8\left(\frac{\psi\Omega_{h}}
{2\cdot10^{-4}}\right)
\left(\frac{\sigma_e}{200\mbox{~km~s}^{-1}}\right)^{(3.66\div4.03)}M_{\odot},
\end{equation}
where pair of coefficients 0.91 and 3.66 corresponds to
$M=10^{10}M_{\odot}$, and respectively pair 1.03 and 4.03 correspond to
$M=10^ {12}M_{\odot}$. The considered model is in a good agreement with
observation data (\ref{korsig}). The fluctuation spectrum at the galactic
scale, $n\approx-2$, completely defines the power index $\alpha\approx4$
in the relation $M_{BH }\propto\sigma_e^\alpha$. There are definite
astrophysical limitations$^{11}$ on the number and mass of PBHs. We
consider the case $\Omega_h\sim10^{-4}$ in accordance with all the limits.

The necessary requirement of the above model is multiple merging of
PBHs with mass $M_h$ into the one SMBH with mass $M_{BH}$ during
the Hubble time. It is known that for a single
black hole with mass $M_h\ll10^7M_{\odot}$ the dynamical friction in the
galactic halo is ineffective. Nevertheless for the early formed PBHs
it is possible the  process of dark matter secondary accretion$^{12}$.  As a
result the PBHs would be ``enveloped'' by the dark matter halo with a
mass of a typical dwarf galaxy and a steep density profile, $\rho\propto
r^{-9/4}$.  Indeed$^{10}$, the gravitationally bound objects formed at
red-shifts $z_{col}$ from the density fluctuations at the moment of
matter--radiation equality \begin{equation}
\delta_{eq}=\delta_c(1+z_{col})/(1+z_{eq}),
\label{zc}
\end{equation}
where $z_{eq}$ is the red-shift of equality and  $\delta_c=1.686$. In the
uniform Universe the PBH with mass $M_h$ produce this fluctuation inside
the sphere containing the total mass $M=M_h/\delta_{eq}$. We will call
this combined spherical volume ``PBH~$+$halo'' by ``induced halo'' (IH).

The growth of IH terminates at the epoch of nonlinear growth of ambient
density fluctuations with the same mass $M$ as IH but originated
from ordinary cosmological perturbation spectrum $P(k)$ with
the r.m.s. fluctuation $\delta_{eq}^{fl}(M)$ in the mass scale $M$:
\begin{equation}
\delta_{eq}^{fl}(M)=\frac{M_h}{M}.
\label{sig1}
\end{equation}
We use for $P(k)$ and $\delta_{eq}^{fl}(M)$ the known
expression$^{13}$ and
solve numerically the
equation
(\ref{sig1}) relative to independent variable
$M$ with the $M_h$ as parameter. See Fig.~1 for calculated relations for
$M(M_h)$ and  $z_{col}(M_h)$ according to (\ref{zc}). For  $M_h=10^5M_{\odot}$
we find $z_{col}\simeq10$ and $M=2\cdot10^7M_{\odot}$. As a result up to epoch
$z\simeq10$ the PBHs with mass $M_h=10^5M_{\odot}$ had time to capture an
additional mass which $\sim200$  times exceeds the PBH mass.
It is easily verified that for $\Omega_h\sim10^
{-4}$ the contribution IHs
to the total galactic mass is negligible.

So we demonstrate that massive IHs with mass $2\cdot10^7M_{\odot}$ are
formed around the PBHs. These IHs are massive enough to sink down to the
galactic center during the Hubble time under influence of dynamical friction.
Although the fate of nested PBHs inside the central parsec of the host galaxy
is rather uncertain, we will suppose that
multiple PBHs merge into a single SMBH during the Hubble time.
Notice that
dynamical friction must be very effective for merging because the
density of IH $\rho\propto r^{-9/4}$ strongly grows towards the center and
smoothed out only at small distance
from PBH.

Our assumption of multiple merging
of PBHs may be violated in the galaxies of late Hubble types.
The central SMBH masses in Sa, Sb, Sc galaxies are less in a mean than in E and
S0 galaxies. In our model this is related with a
relatively late formation of Sa, Sb, Sc galaxies when the main part of PBHs
do not have enough time to sinking to the galactic center. In particular
$\sim10^2$ PBHs of mass $M_h\sim10^5M_{\odot}$ can inhabit our Galaxy.

Coalescence of PBHs in the galaxies must be accompanied by the strong
burst of gravitational radiation. The projected interferometric detector
LISA is capable to detect this coalescence. So there is a principal
possibility for the verification of considered model by the LISA detector.

In conclusion we discuss why the PBHs with masses $\sim10^5M_{\odot}$ are
very probable candidates on the independent black holes population.
Noncompact objects (neutralino stars) of mass  $\sim(0.1\div1)M_{\odot}$,
consisting of weakly interacting nonbaryonic dark matter particles
like neutralino were proposed by Gurevich et
al.$^{14}$  for the explanation of microlensing events in Large
Magellanic Clouds. The hypothesized neutralino stars are originated from
the cosmological fluctuations with a narrow sharp maximum $\sim1$ in the
spectrum at some small scale. In addition to neutralino stars the same
maximum in the spectrum of cosmological fluctuations produce$^{15}$ also
the massive PBHs with mass $\sim10^5M_{\odot}$. So the hypothesized dark
matter neutralino stars and PBHs may be indirectly connected through
their common origin from the same cosmological fluctuations. The spectrum
with a sharp maximum at some scale arises in some inflation models for
example in the model of Starobinsky$^{16}$. At the same time the spectrum
beyond the maximum may be of the standard Harrison-Zel'dovich form and
reproduce the usual scenario of large-scale structure formation in the
galactic distribution.

The work was supported in part by the INTAS grant 99--1065 and by Russian
Foundation for Basic Research grants 01-02-17829, 00-15-96697 and
00-15-96632.

\newpage

\begin{figure}
\resizebox{\hsize}{!}{\includegraphics{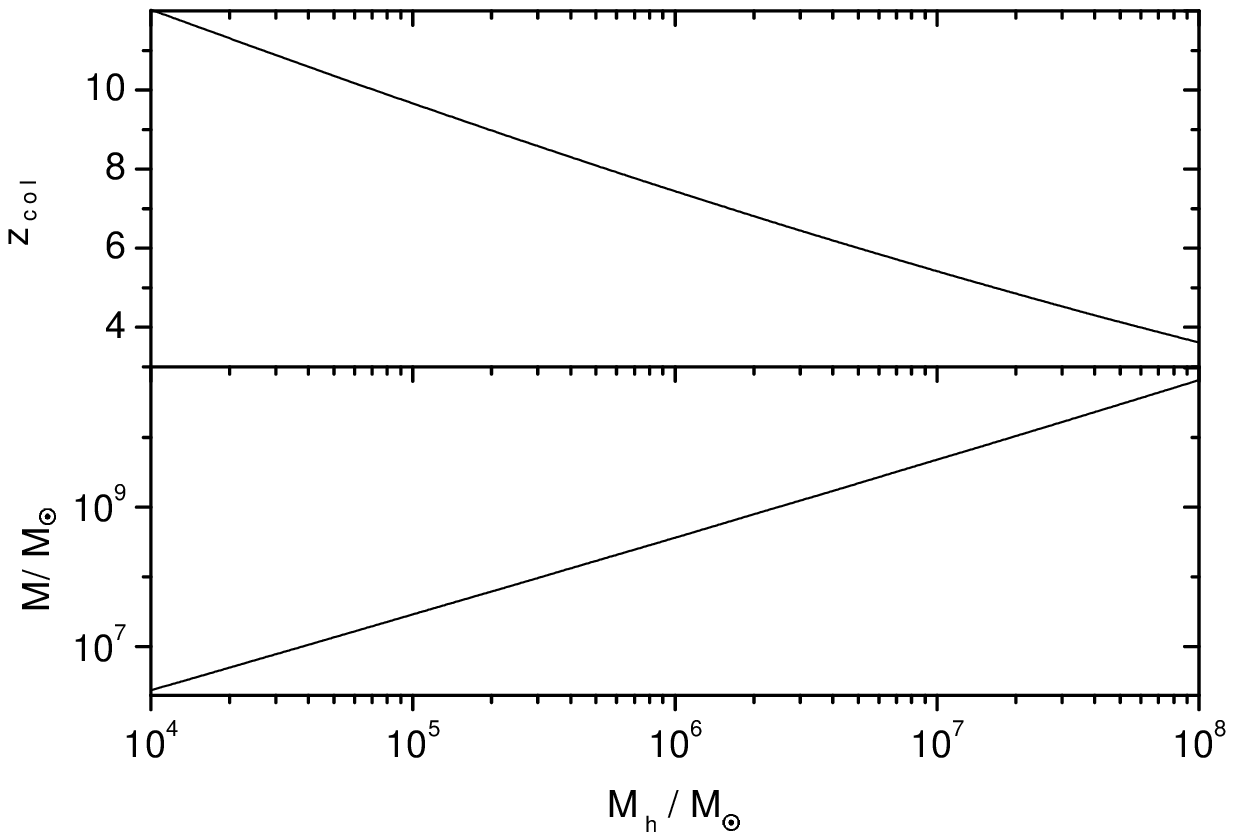}} 
\caption{ Red-shift of induced halo formation $z_{col}(M_h)$ as a
function of its mass $M(M_h)$ from the numerical solution of the equation
(\ref{sig1}). } \label{fig1}
\end{figure}


\begin{thebibliography}{99}


\bibitem{1}Kormendy, J. and Richstone, D.,
1995, {\it ARA\&A} {\bf 33}, 581


\bibitem{2}Gebhardt, K. et al.,
2000, {\it ApJ} {\bf 539}, L642

\bibitem{3}Ferrarese, L. and Merritt, D., 2000, {\it ApJ} {\bf 539}, L9;
astro-ph/0006053

\bibitem{4}Silk, J. and Rees, M.J., 1998, astro-ph/9801013

\bibitem{5}Fukugita, M. and Turner, E.L.,
1996, {\it ApJ} {\bf 460}, L81

\bibitem{6}Rees, M. J. 1984,
{\it ARA\&A}, {\bf 22}, 471


\bibitem{7}Zel'dovich, Ya.B. and Novikov, I.D.,
1967, Sov. Astron. {\bf 10},  602

\bibitem{8}Carr, B.J., 1975,
{\it ApJ} {\bf 201}, 1

\bibitem{9}Rubin, S.G., Khlopov, M.Yu. and Sakharov, A.S., 2001, {\it JETP}
92, 921; hep-ph/0106187


\bibitem{10}White, S.D.M, 1994, astro-ph/9410043

\bibitem{11}Dokuchaev, V.I. and Eroshenko, Yu.N., 2001, {\it Astronomy
Letters} {\bf 27}, 759, astro-ph/0202019

\bibitem{12}Gunn, J.E., 1977,
{\it ApJ} {\bf 218}, 592

\bibitem{13}Barden, J.M. et al., 1986,
{\it ApJ} {\bf 304}, 15

\bibitem{14}Gurevich, A.V., Zybin, K.P. and Sirota, V.A., 1997, {\it Sov.
Phys. Usp.} {\bf 167}, 913, astro-ph/9801314


\bibitem{15}Dokuchaev, V.I. and Eroshenko, Yu.N., 2001, {\it JETP} {\bf 94},
5, astro-ph/0202021


\bibitem{16}Starobinsky, A.A., 1992, {\it JETP Lett.}  {\bf 55}, 489

\end{thebibliography}
\end{document}